\documentclass[aps,onecolumn,groupedaddress,nofootinbib]{revtex4}
\usepackage{dsfont}
\usepackage{bm}

\begin{document}

\title{Lifshitz black brane thermodynamics in the presence of a nonlinear
electromagnetic field}
\author{M. H. Dehghani$^{1,2}$ \footnote{%
email address: mhd@shirazu.ac.ir}, Ch. Shakuri$^2$ and M. H. Vahidinia$^2$ \footnote{%
email address: vahidinia@shirazu.ac.ir}}
\affiliation{$^1$ Center for Excellence in Astronomy \& Astrophysics (CEAA -- RIAAM),
Maragha, IRAN, P. O. Box: 55134 - 441}
\affiliation{$^2$Physics Department and Biruni Observatory, College of Sciences, Shiraz
University, Shiraz 71454, Iran}

\begin{abstract}
In this paper, we investigate the thermodynamics of Lifshitz black branes
in the presence of a nonlinear massless electromagnetic field. We begin by introducing the
appropriate action in grand-canonical and canonical ensembles for nonlinear
electromagnetic field. The condition on the parameters
of the metric for having black brane solutions will be presented.  Since the field equations cannot be solved for an arbitrary
value of the critical exponent $z$, we obtain a conserved quantity along the $r$ coordinate that
enables us to relate the parameters of the metric at the horizon and at infinity. Then,
we calculate the energy density of the Lifshitz black brane through the use of the counterterm
method generalized for the asymptotic Lifshitz spacetimes. Finally, we present a relation
between the energy density and the thermodynamical quantities, electric potential, charge density,
temperature and entropy density. This relation is the generalization of Smarr formula
for anti-de Sitter black branes.
\end{abstract}
\pacs{04.70.Bw, 04.70.Dy, 04.50.Gh, 11.25.Tq}
\maketitle

\section{Introduction}

Quantum field theories can describe condensed matter systems at distance
scales large compared to the lattice spacing \cite{Alt}. For these systems,
when the underlying electronic ground state undergoes a continuous
nonanalytic change as a function of an external parameter, such as pressure,
the resulting quantum field theory becomes quantum critical \cite{Sach}. The
quantum critical theories arising in condensed matter systems are scale
invariant, but, in general, space and time need not scale equally ($%
t\rightarrow \lambda ^{z}t$, $x\rightarrow \lambda \mathbf{x}$, $z\neq 1$)
\cite{Hertz}. Thus, the holographic description of field theories with anisotropic
scaling symmetry presents an interesting extension of AdS/CFT, which may
have valuable applications in condensed matter theory. The simplest example
of such a dual description is the Lifshitz metric
\begin{equation}
ds^{2}=-\frac{r^{2z}}{l^{2z}}dt^{2}+\frac{l^{2}}{r^{2}}dr^{2}+r^{2}d\mathbf{x}%
^{2}.  \label{lif}
\end{equation}%
originally constructed in Ref. \cite{Kach}. Understanding such symmetries
holographically requires us to go beyond the familiar context of
asymptotically anti-de Sitter spacetimes on the gravitational side. This
duality is interesting both for its potential application in condensed
matter, and as an extension of our understanding of holography and the
relation between field theory and spacetime descriptions. The use of
gravitational duals to study strongly coupled field theories \cite{Mal} has
provided a unique tool that has shed light on a number of important
questions, particularly concerning thermodynamics and theories at finite
density, which are related to black holes in the bulk. This work has been
extended over the last few years to include applications to field theories
of interest to condensed matter physics (see Ref. \cite{Hart} for useful
reviews). Indeed, the counterterm method for Lifshitz black holes in the
presence of a massive electromagnetic field has been introduced in \cite{omid}.
Here, we want to use the counterterm method introduced in Ref. \cite{omid}
in the case of an asymptotically Lifshitz black brane in
the presence of a nonlinear electromagnetic field.

The action of Einstein-Hilbert (EH) does not admit Lifshitz geometry. But by adding a nontrivial matter field to the EH action, the new action may admit Lifshitz geometry \cite{Kach}. Also, higher curvature terms
in the action may produce anisotropic scaling symmetry \cite{Pang}. Although
the field equations of the EH action in the presence of a nontrivial matter
field such as a massive electromagnetic field, a dilaton field coupled with an electromagnetic field, or other more complicated
matter field are not
easy to solve, a few analytical solutions of asymptotic Lifshitz
spacetime have been obtained in Ref. \cite{Exact}. Numerical methods have been also
employed to study a continuous range of $z$ for both black holes and black
branes \cite{Num}. Also, the thermodynamics of
Lifshitz black holes in various theories of gravity in the presence of a
massive gauge field have been investigated \cite{Therm,MH,Pour}.

Here, we add a power-law Maxwell invariant term to the action of EH
in the presence of a massive linear electromagnetic field. This higher term, $%
(-F_{\mu \nu }F^{\mu \nu })^{s}$, also appears in the low-energy limit of
heterotic string theory \cite{Gross}. Also, it is worth noting that this
term for $s=d/4$, is conformally invariant \cite{Mart}. That is, this Lagrangian $(-F_{\mu \nu }F^{\mu \nu })^{d/4}$, is
invariant under the conformal transformation $g_{\mu \nu }\rightarrow \Omega
^{2}g_{\mu \nu }$ and $A_{\mu }\rightarrow A_{\mu }$. The black hole
solutions of various theories of gravity in the presence of this power-law
Lagrangian have been investigated by many authors \cite{Hendi}.

In this paper, we like to investigate the thermodynamics of Lifshitz black branes in
the presence of a nonlinear massless gauge field. In order to do this, we should first introduce the appropriate
action of the system in canonical and grand-canonical ensembles. The appropriate action of
Einstein gravity both in canonical and grand-canonical ensembles
in the presence of a linear gauge field has been introduced by Hawking and Ross \cite{Haw}. This action has been used
by many authors in investigating the thermodynamics of charged black holes \cite{Can}. Our first aim is to generalize this
action for linear electromagnetic field to the case of nonlinear gauge field both in canonical and grand-canonical ensembles.
Having the appropriate action
in grand-canonical and canonical ensembles, we use the counterterm method to calculate the
finite energy density of the black brane solutions.
We also generalize the Smarr formula of black holes to the case of charged-Lifshitz black branes.

The outline of our paper is as follows. In Sec. \ref{Field}, we give a brief
review of the Einstein equation in the presence of a massive and a nonlinear
massless electromagnetic fields. In Sec. \ref{Act}, we introduce the action of
the nonlinear massless gauge field appropriate for grand-canonical and canonical ensembles.
We also review the finite action  and finite energy density of the theory.
Section \ref{Black} is devoted to the investigation of the existence of black brane solutions.
Since presenting an exact solution is not easy, we introduce a conserved quantity along
the $r$ coordinate in Sec \ref{C0}, in order to investigate the thermodynamics of the black brane solutions.
In Sec. \ref{Therm}, we consider the thermodynamic properties of charged Lifshitz black branes, and present
the energy density in terms of the thermodynamic quantities of the solutions. Finally, we finish our paper with some
concluding remarks.

\section{Field Equations \label{Field}}

The bulk action of Einstein gravity in the presence of two Abelian massless $%
A_{\mu }$ and massive $B_{\mu }$ vector fields in $(n+1)$ dimensions may be
written as

\begin{eqnarray}
I_{\mathrm{bulk}} &=&\frac{1}{16\pi }\int_{\mathcal{M}}d^{n+1}x\sqrt{-g}%
\left( \mathcal{L}_{g}+\mathcal{L}_{m}\right) ,  \nonumber \\
\mathcal{L}_{g} &=&R-2\Lambda ,  \nonumber \\
\mathcal{L}_{m} &=&\frac{1}{4}(-F)^{s}-\frac{1}{2}CB_{\mu }B^{\mu }-\frac{1}{%
4}H_{\mu \nu }H^{\mu \nu },  \label{Act1}
\end{eqnarray}%
where $\Lambda $ is the cosmological constant, $R$ is the Ricci scalar, $F=F_{\mu
\nu }F^{\mu \nu }$ is the Maxwell invariant and $C$ is the mass of
the massive electromagnetic field. $F_{\mu \nu }=\partial _{\lbrack \mu }A_{\nu ]}$ and $%
H_{\mu \nu }=\partial _{\lbrack \mu }B_{\nu ]}$ are electromagnetic tensor
fields related to massless and massive vector fields $A_{\mu }$ and $B_{\mu }$%
, respectively . The minus sign coupled with $F$ in the Lagrangian of the
nonlinear electromagnetic field is due to the fact that the Maxwell
invariant is negative. For $s=0$, the first term in $\mathcal{L}_{m}$ would
be a constant that can be absorbed in the cosmological constant; therefore
we ignore this case.

The variation of the action (\ref{Act1}) with respect to the gravitational field
$g_{\mu \nu }$ and the vector fields $A_{\mu }$ and $B_{\mu }$ yields
\begin{eqnarray}
&& \nabla ^{\mu }H_{\mu \nu }=CB_{\mu },  \label{EqH1}\\
&& \partial _{\mu }\left( \sqrt{-g}F^{\mu \nu }(-F)^{s-1}\right) =0,
\label{EqF} \\
&& G_{\mu \nu }+\Lambda g_{\mu \nu } =T_{\mu \nu },  \label{EqG}
\end{eqnarray}
where the energy-momentum tensor $T_{\mu \nu }$ is

\begin{equation}
T_{\mu \nu }=\frac{1}{2}\left[ s(-F)^{s-1}F_{\mu \rho }F_{\nu }^{%
\phantom{\nu}{\rho}}+\frac{1}{4}(-F)^{s}g_{\mu \nu }\right] +\frac{1}{2}%
(H_{\phantom{\rho}{\mu}}^{\rho }H_{\rho \nu }-\frac{1}{4}H_{\rho \sigma
}H^{\rho \sigma }g_{\mu \nu }+C[B_{\mu }B_{\nu }-\frac{1}{2}B_{\rho }B^{\rho
}g_{\mu \nu }]).  \label{Tmn}
\end{equation}
In the limit of $s=1$, the nonlinear electromagnetic field equation reduces
to the standard form of Maxwell equation.

Here, we like to consider the thermodynamics of the asymptotic Lifshitz black
hole. The metric of an $(n+1)$-dimensional asymptotically Lifshitz static
spacetime with zero curvature boundary may be written as

\begin{eqnarray}
ds^{2}
&=&-e^{2F(r)}dt^{2}+e^{2G(r)}dr^{2}+l^{2}e^{2R(r)}\sum%
\limits_{i=1}^{n-1}(dx^{i})^{2}  \label{met1} \\
&=&-\frac{r^{2z}}{l^{2z}}f(r)dt^{2}+\frac{l^{2}dr^{2}}{r^{2}g(r)}%
+r^{2}\sum\limits_{i=1}^{n-1}(dx^{i})^{2},  \label{met2}
\end{eqnarray}
where the metric functions are related through the following relations

\begin{eqnarray}
F(r) &=&\ln \left( \frac{r^{z}}{l^{z}}\sqrt{f(r)}\right) ,  \nonumber \\
R(r) &=&\ln \left( \frac{r}{l}\right) ,  \nonumber \\
G(r) &=&-\ln \left( \frac{r}{l}\sqrt{g(r)}\right).  \label{func}
\end{eqnarray}

Using the Ansatz
\begin{equation}
A=qe^{K(r)}dt=\frac{q}{l^{z}}k(r)dt  \label{Aeq}
\end{equation}
for the massless gauge field, Eq. (\ref{EqF}) reduces to
\begin{equation}
(2s-1)rfgk^{^{\prime \prime }}+\frac{1}{2}k^{^{\prime
}}\{-(2s-1)rgf^{^{\prime }}+f[(2s-1)rg^{^{\prime }}-2g[2(z-1)s-z-(n-2)]]\}=0
\label{NMF}
\end{equation}
which indicates that the nonlinear electromagnetic field vanishes for $s=1/2$%
. The solution of Eq. (\ref{NMF}) may be written as

\begin{equation}
k^{\prime }(r)=r^{-(m+1-z)}\sqrt{\frac{f}{g}},  \label{K'}
\end{equation}%
where $m=(n-1)/(2s-1)$. In this paper, we choose the horizon as the reference
point of the potential $A_{\mu}$. Thus, one obtains
\begin{eqnarray*}
k(r) &=&\int^{r}r^{-(m+1-z)}\sqrt{\frac{f}{g}}dr+D, \\
D &=&-\int^{r_{0}}r^{-(m+1-z)}\sqrt{\frac{f}{g}}dr,
\end{eqnarray*}%
where $r_{0}$ is the horizon radius. One may note that $D$ is positive, and therefore
the potential is negative in the cases that the first integral vanishes at infinity.

Using Eq. (\ref{K'}) and the Ansatz
\begin{equation}
B=Qe^{H(r)}dt=Q\frac{r^{z}}{l^{z}}h(r)dt  \label{Beq}
\end{equation}
for the massive vector field, the equations of motion (\ref{EqH1}) and (\ref%
{EqG}) reduce to the following system of nonlinear coupled differential
equations:
\begin{eqnarray}
2r^{2}h^{\prime \prime }-r\left[ (\ln f)^{\prime }-(\ln g)^{\prime }\right]
(rh^{\prime }+z)-2(n+z)rh^{\prime }+2(n-1)z &=&\frac{2Cl^{2}}{g},  \nonumber
\\
n(n-1)r^{2}g+(n-1)r^{3}g^{\prime }+2\Lambda
l^{2}r^{2}-2l^{2}r^{2}T_{t(massive)}^{t} &=&2l^{2}r^{2}T_{t(massless)}^{t},
\nonumber \\
(n-1)(n-2+2z)r^{2}g+(n-1)r^{3}g(\ln f)^{\prime }+2\Lambda
l^{2}r^{2}-2l^{2}r^{2}T_{r(massive)}^{r} &=&2l^{2}r^{2}T_{r(massless)}^{r},
\label{Fieldeq}
\end{eqnarray}
where prime denotes the derivative with respect to $r$. In the above
equations, the component of the energy momentum tensors are

\begin{eqnarray*}
T_{t(massive)}^{t} &=&-\frac{Q^{2}}{4l^{2}f}\left\{ g(rh^{\prime
}+zh)^{2}+Cl^{2}h^{2}\right\} , \\
T_{r(massive)}^{r} &=&-\frac{Q^{2}}{4l^{2}f}\left\{ g(rh^{\prime
}+zh)^{2}-Cl^{2}h^{2}\right\} , \\
T_{t(massless)}^{t} &=&-\frac{(2s-1)}{8}\left( \frac{2q^{2}}{l^{2}r^{2m}%
}\right) ^{s}, \\
T_{r(massless)}^{r} &=&-\frac{(2s-1)}{8}\left( \frac{2q^{2}}{l^{2}r^{2m}%
}\right) ^{s}.
\end{eqnarray*}

In order to have asymptotically Lifshitz geometry, the field equations (\ref%
{Fieldeq}) should be satisfied for $f(r)=g(r)=h(r)=1$ and $k(r)=D$ as $%
r\rightarrow \infty $. Then, we obtain the following constraints on $C$, $Q$
and $\Lambda $:

\begin{eqnarray}
C &=&\frac{(n-1)z}{l^{2}},\qquad Q^{2}=\frac{2(z-1)}{z},  \nonumber \\
\Lambda &=&-\frac{(z-1)^{2}+n(z-2)+n^{2}}{2l^{2}}.  \label{cons}
\end{eqnarray}

\section{Finite Action in Canonical and Grand-Canonical Ensembles \label{Act}}

As it is known, the bulk action of gravity neither has a well-defined
variational principle nor is finite. In order to have a finite action with a
well-defined variational principle, one should add some boundary terms to
it. The boundary terms for the action (\ref{Act1}) without the massless gauge
field $A^{\mu }$ may be written as \cite{omid}

\begin{equation}
I_{bdy}=\frac{1}{8\pi }\int_{\partial \mathcal{M}}d^{n}x\sqrt{-h}\left[ K-%
\frac{n-1}{l}-\frac{zQ}{2l}(-B_{\gamma }B^{\gamma })^{1/2}\right] +I_{deriv},
\label{ActB}
\end{equation}%
where $\partial \mathcal{M}$ is the hypersurface at some constant $r$, $%
h_{\alpha \beta }$ is the induced metric, $K$ is the trace of the extrinsic
curvature $K_{\alpha \beta }=\nabla _{(\alpha }n_{\beta )}$ of the boundary
and the unit vector $n^{\mu }$ is orthogonal to the boundary and outward
directed. The last term $I_{deriv}$ is a collection of terms involving
derivatives of the boundary fields that could involve both the curvature
tensor constructed from the boundary metric and covariant derivatives of $%
B_{\alpha }$. Since the boundary is flat and the fields are constants, this
term will not contribute to the on-shell action for the Lifshitz solution or its
first variation around the Lifshitz background and, therefore, we ignore it.

The first term in the action (\ref{ActB}) makes the
variational principle of the EH action well-defined, while the second and third terms are added
to the bulk action in order to cancel the divergences of the gravity action
and that of the massive vector field \cite{omid}. The variation of the total
action ($I=I_{bulk}+I_{bdy}$) about the solutions of the equations of motion
is

\begin{equation}
\delta I=\int d^{n}x(S_{\alpha \beta }\delta h^{\alpha \beta }+S_{\alpha
}^{L}\delta B^{\alpha })+\frac{1}{16\pi }\int_{\partial \mathcal{M}}d^{n}x%
\sqrt{-h}s(-F)^{s-1}n^{\mu }F_{\mu \alpha }\delta A^{\alpha },
\label{delta-I}
\end{equation}
where

\begin{eqnarray}
&&S_{\alpha \beta }=\frac{\sqrt{-h}}{16\pi }\left[ \Pi _{\alpha \beta }+%
\frac{zQ}{2l}(-B_{\gamma }B^{\gamma })^{-1/2}(B_{\alpha }B_{\beta
}-B_{\gamma }B^{\gamma }h_{\alpha \beta })\right] , \\
&&S_{\alpha }^{L}=-\frac{\sqrt{-h}}{16\pi }\left[ n^{\mu }H_{\mu \alpha }-%
\frac{zQ}{l}(-B_{\gamma }B^{\gamma })^{-1/2}B_{\alpha }\right] ,
\end{eqnarray}

\begin{equation}
\Pi _{\alpha \beta }=K_{\alpha \beta }-Kh_{\alpha \beta }+\frac{(n-1)}{l}%
h_{\alpha \beta }.
\end{equation}%
Thus, the variation of the total action with respect to $A^{\mu }$ will only
give the equation of motion of the nonlinear massless field $A^{\mu }$, if the
variation is at fixed nonlinear massless gauge potential on the boundary. That is, the
total action, $I=I_{bulk}+I_{bdy}$ given in Eqs. (\ref{Act1}) and
(\ref{ActB}) is appropriate for the grand-canonical ensemble, where $\delta
A^{\mu }=0$ on the boundary. But in the canonical ensemble, where the
electric charge $[s(-F)^{s-1}n^{\mu }F_{\mu \alpha }]$ is fixed on the
boundary, the appropriate action is
\begin{equation}
I=I_{bulk}+I_{bdy}-\frac{1}{16\pi }\int_{\partial \mathcal{M}}d^{n}x\sqrt{-h}%
s(-F)^{s-1}n^{\mu }F_{\mu \alpha }A^{\alpha }.  \label{ActCan}
\end{equation}%
The last term in Eq. (\ref{ActCan}) is the generalization of the boundary
term of the linear electromagnetic field introduced by Hawking and Ross \cite{Haw} to
the case of the nonlinear electromagnetic field with a power law Lagrangian. Of
course, one may note that this boundary term reduces to the boundary term of
Hawking and Ross for $s=1$. Thus, both in the canonical and grand-canonical ensemble, the
variation of total action about the solutions of the field equations is
\[
\delta I=\int d^{n}x(S_{\alpha \beta }\delta h^{\alpha \beta }+S_{\alpha
}^{L}\delta B^{\alpha }).
\]%
That is, the nonlinear gauge field is absent in the variation of the total
action both in canonical and grand-canonical ensembles, and, therefore, as in
the absence of a massless electromagnetic field, the dual field theory for
asymptotically Lifshitz spacetime in the presence of the nonlinear
electromagnetic field has a stress tensor complex consisting of the energy
density $\mathcal{E}$, energy flux $\mathcal{E}_{i}$, momentum density $%
\mathcal{P}_{i}$, and spatial stress tensor $\mathcal{P}_{ij}$ satisfying the
conservation equations \cite{omid}

\begin{equation}
\partial _{t}\mathcal{E}+\partial _{i}\mathcal{E}^{i}=0,\qquad \partial _{t}%
\mathcal{P}_{j}+\partial _{i}\mathcal{P}_{\phantom{i}{j}}^{i}=0,
\end{equation}%
where
\begin{eqnarray}
\mathcal{E} &=&2S_{\phantom{t}{t}}^{t}-S_{L}^{t}B_{t},\qquad \mathcal{E}%
^{i}=2S_{\phantom{i}{t}}^{i}-S_{L}^{i}B_{t},  \label{En} \\
\mathcal{P}_{i} &=&-2S_{\phantom{t}{i}}^{t}+S_{L}^{t}B_{i},\qquad \mathcal{P}%
_{\phantom{j}{i}}^{j}=-2S_{\phantom{j}{i}}^{j}+S_{L}^{j}B_{i}.  \label{Mom}
\end{eqnarray}

\section{Black Brane Solutions \label{Black}}

In order to investigate the black brane solutions, we may consider the near-horizon behavior.
The near-horizon behavior of the functions $f(r)$, $g(r)$%
, $h(r)$ and $k(r)$ may be written as
\begin{eqnarray}
f(r) &=&f_{1}\left\{
(r-r_{0})+f_{2}(r-r_{0})^{2}+f_{3}(r-r_{0})^{3}+f_{4}(r-r_{0})^{4}+...\right%
\} ,  \nonumber \\
g(r)
&=&g_{1}(r-r_{0})+g_{2}(r-r_{0})^{2}+g_{3}(r-r_{0})^{3}+g_{4}(r-r_{0})^{4}+...,
\nonumber \\
h(r) &=&f_{1}^{1/2}\left\{
h_{0}+h_{1}(r-r_{0})+h_{2}(r-r_{0})^{2}+h_{3}(r-r_{0})^{3}+h_{4}(r-r_{0})^{4}+...\right\} ,
\nonumber \\
k(r) &=&f_{1}^{1/2}\left\{
k_{1}(r-r_{0})+k_{2}(r-r_{0})^{2}+k_{3}(r-r_{0})^{3}+k_{4}(r-r_{0})^{4}+...%
\right\}   \label{ExpH}
\end{eqnarray}%
where $f_{i}$'s, $g_{i}$'s, $h_{i}$'s and $k_{i}$'s are constants. One may
note that for a nonextreme black brane, both $f_{1}$ and $g_{1}$ should be
nonzero. If $g_{1}=0$, then one may have an extreme black brane. Inserting
these expansions into the equation of motion and demanding that the coefficients
for each power of $(r-r_{0})$ vanish determines all the coefficients in terms of $%
r_{0}$. One finds that $h_{0}=0$, and
\[
g_{1}=z\frac{z^{2}+z(n-2)+(n-1)^{2}+2l^{2}T_{r(massless)}^{r}(r=r_{0})}{%
z(n-1)r_{0}+(z-1)r_{0}^{2}h_{1}^{2}}.
\]
Now, in order to have nonextreme black brane solution, $g_{1}$ should be
positive, and, therefore,
\begin{equation}
\left[ z^{2}+z(n-2)+(n-1)^{2}\right] r_{0}^{2sm}>\frac{(2s-1)l^{2}}{4}%
\left( \frac{2q^{2}}{l^{2}}\right) ^{s}.  \label{con1}
\end{equation}

\section{A conserved quantity along the radial coordinate \label{C0}}

Exact solutions of the field equations (\ref{Fieldeq}) for continuous values
of $z$ are not easily obtained. In order to investigate the
thermodynamics of Lifshitz black branes in the presence of a nonlinear gauge
field, we introduce a conserved quantity along the coordinate $r$. This
conserved quantity can relate the quantities on the horizon such as
temperature and entropy, and those at infinity, such as energy density.
Inserting the Ansatze (\ref{Aeq}) and (\ref{Beq}) into the action (\ref{Act1}%
) and integrating by part, one obtains the one-dimensional Lagrangian as
\begin{eqnarray}
\mathcal{L}_{1D} &=&(n-1)l^{n-1}\left\{ -2\frac{\Lambda }{n-1}e^{2G}+\left[
2F^{\prime }R^{\prime }+(n-2)R^{\prime 2}\right] \right\} e^{F-G+(n-1)R}
\nonumber \\
&&+\frac{l^{n-1}}{2}Q^{2}\left( C+H^{\prime 2}e^{-2G}\right)
e^{-F+G+(n-1)R+2H}+\frac{1}{4}\left( \frac{2q^{2}K^{\prime 2}e^{2K}}{%
e^{2F+2G}}\right) ^{s}e^{F+G+(n-1)R}.  \label{Act2}
\end{eqnarray}
Using the same procedure discussed in Ref. \cite{MH}, one obtains the conserved
quantity along the radial coordinate $r$ as
\begin{eqnarray}
\mathcal{C}_{0} &=&\left\{ \left[ rf^{\prime }+2(z-1)f\right]
-Q^{2}(zh+rh^{\prime })h\right\} \frac{r^{n+z-1}}{l^{z+1}}\sqrt{\frac{g}{f}}
\nonumber \\
&&-2^{s-1}s(\frac{q^{2}r^{2}k^{2}g}{r^{2z}l^{2}f})^{s}(\frac{k^{\prime }}{k}%
)^{2s-1}\frac{r^{n+z-2}}{l^{z-1}}\sqrt{\frac{f}{g}}  \label{Cons}
\end{eqnarray}%
which can be written as

\[
\mathcal{C}_{0}=(rf^{\prime }+2(z-1)f-Q^{2}(zh+rh^{\prime })h)\frac{r^{n+z-1}%
}{l^{z+1}}\sqrt{\frac{g}{f}}-\frac{2^{s-1}sq^{2s}}{l^{z-1+2s}}k.
\]

In the case of $z=1$ ($Q=0$), $f(r)=g(r)$ and, therefore, the constant $\mathcal{C}_{0}$
reduces to

\[
\mathcal{C}_{0}=\frac{r^{n+1}}{l^{2}}f^{\prime }+\frac{2^{s-1}sq^{2s}}{l^{2s}%
}\left( \frac{2s-1}{(n-2s)r^{(n-2s)/(2s-1)}}-D\right),
\]%
which gives the following solution for the metric function

\[
\frac{r^{2}}{l^{2}}f(r)=\frac{r^{2}}{l^{2}}-\frac{m}{r^{n-2}}+\frac{%
(2s-1)^{2}2^{s-1}q^{2s}}{(n-2s)(n-1)l^{2s}r^{2[(n-3)s+1]/(2s-1)}},
\]%
where $m=(\mathcal{C}_{0}+2^{s-1}sq^{2s}l^{-2s}D)/n$. This is the asymptotic
AdS solution introduced in Ref. \cite{Mart}.

In order to relate the parameters of the metric function on the horizon and
infinity, we calculate the constant $\mathcal{C}_{0}$ on the
horizon at infinity in terms of the thermodynamic quantities. Using the near horizon expansions presented in the last section, one can
show that
\begin{equation}
\mathcal{C}_{0}=\frac{\sqrt{f_{1}g_{1}}}{l^{z+1}}r_{0}^{(n+z)}.  \label{Chor}
\end{equation}%

The constant $\mathcal{C}_{0}$ at infinity can be calculated by using the large $r$ expansions given in the appendix as
\begin{equation}
\mathcal{C}_{0}=\frac{2(z-1)(z-n+1)(z+n-1)^{2}C_{1}}{zl^{(z+1)}\mathcal{K}}-\frac{2^{s-1}q^{2s}k_{\infty }}{l^{z+2s-1}},  \label{C}
\end{equation}%
where $k_{\infty }=k(r\rightarrow \infty )$ and $\mathcal{K}$ is
\[
\mathcal{K}=(z-1)(n+z-1)+z(z-1)+n(n-1).
\]
The constant $C_{0}$ is finite at the horizon and, therefore, it should be
finite at infinity too. Thus, one has some restrictions on the values of $n$%
, $s$ and $z$. For positive $z$,\ $k_{\infty }$ is finite, provided:

\textbf{A}. $s>1/2$ and $z<(n-1)/(2s-1)$: In this case $\mathcal{C}_{0}$ is
\[
\mathcal{C}_{0}=\frac{2(z-1)(z-n+1)(z+n-1)^{2}C_{1}}{zl^{(z+1)}\mathcal{K}}-%
\frac{2^{s-1}q^{2s}D}{l^{z+2s-1}},
\]

\textbf{B. }$z=n-1$ and $1/2<s<1$: For this case, one obtains
\begin{equation}
\mathcal{C}_{0}=\frac{2C_{1}}{(2n-3)l^{z+1}}-\frac{2^{s-1}q^{2s}D}{l^{z+2s-1}%
}.  \label{CC}
\end{equation}%
Now, the relation between the parameters of the functions $f$, $g$ and $k$
at the horizon and infinity is
\begin{eqnarray*}
\frac{\sqrt{f_{1}g_{1}}}{l^{z+1}}r_{0}^{(n+z)} &=&\frac{%
2(z-1)(n-z-1)(z+n-1)^{2}C_{1}}{zl^{z+1}\mathcal{K} }-\frac{2^{s-1}q^{2s}D}{%
l^{2s+z-1}}, \\
\frac{\sqrt{f_{1}g_{1}}}{l^{z+1}}r_{0}^{(n+z)} &=&\frac{2C_{1}}{(2n-3)l^n%
}-\frac{2^{s-1}q^{2s}D}{l^{n+2(s-1)}},
\end{eqnarray*}%
for the cases a and b, respectively. Here, again, one can present the condition of having
nonextreme black branes as:
\[ C_1>\frac{2^{s-2}q^{2s}zD\mathcal{K}}{(z-1)(n-z-1)(n+z-1)^2l^{2s-2}},\]
\[ C_1>\frac{2^{2s-2}(2n-3)q^{2s}D}{l^{2s-2}},\]
for the cases a and b, respectively. The above conditions are in terms of $q$ and the parameters
of the metric at infinity, which are $C_1$ and $D$. However, the condition presented
in Eq. (\ref{con1}) was in terms of $q$ and $r_0$.
\section{Thermodynamics of charged Lifshitz black branes \label{Therm}}

Now, we investigate the thermodynamics of charged Lifshitz black brane
solutions. The entropy per unit volume of the horizon is
\begin{equation}
\mathcal{S}=\frac{r_{0}^{n-1}}{4}.  \label{Ent}
\end{equation}%
One can obtain the temperature of the event horizon by using standard the
Wick-rotation method, yielding the result
\begin{equation}
T=\frac{r_{0}^{z+1}}{4\pi l^{z+1}}\left( f^{\prime }g^{\prime }\right)
_{r=r_{0}}^{1/2}=\frac{\sqrt{f_{1}g_{1}}r_{0}^{z+1}}{4\pi l^{z+1}}.
\label{Temp}
\end{equation}

Using Eq. (\ref{En}), one obtains the energy density of the black brane as
\begin{equation}
\mathcal{E}=\frac{1}{16\pi }\frac{r^{n+z-1}}{l^{z+1}}\left\{2(n-1)\sqrt{f}%
(1-\sqrt{g})+Q^{2}zh \left(1-h\sqrt{\frac{g}{f}}\right)-rQ^{2}\sqrt{\frac{g}{f}}%
hh^{\prime }\right\}.  \nonumber
\end{equation}%
Inserting the large $r$ expansions given in the appendix for the metric
function in the above equation, the energy density for the two cases A and B
discussed in the previous section may be calculated as:
\begin{eqnarray}
\mathcal{E} &=&\frac{(n-1)(z-1)(n+z-1)(n-z-1)}{8\pi z\mathcal{K}l^{z+1}}%
C_{1},  \label{EnA} \\
\mathcal{E} &=&\frac{C_{1}}{16\pi (2n-3)l^{z+1}},
\end{eqnarray}%
respectively.

Now, using the definition of electric potential at infinity with
respect to the horizon \cite{Can}
\[
\Phi =A_{\mu }\chi ^{\mu }\left\vert _{r\rightarrow \infty }-A_{\mu }\chi
^{\mu }\right\vert _{r=r_{0}},  \nonumber
\]
where $\chi ^{\mu }=\partial /\partial t$ is the null generators of the
event horizon, the electric potential is obtained as%
\begin{equation}
\Phi =\frac{q}{l^{z}}D.
\end{equation}

The electric charge density may be calculated by using
\[
\mathcal{Q}=\frac{1}{16\pi \Omega _{k}}\int d\Omega _{k}r^{n-1}n^{\mu
}(-F)^{s-1}F_{\mu \nu }u^{\nu },
\]%
where $n^{\mu }$ and $u^{\nu }$are the unit spacelike and timelike unit
normals to a sphere of radius $r$,
\[
u^{\nu }=\frac{1}{\sqrt{-g_{tt}}}dt=\frac{l^{z}}{r^{z}\sqrt{f}}dt,\text{ \ }%
n^{\mu }=\frac{1}{\sqrt{g_{rr}}}dr=\frac{r\sqrt{g}}{l}dr.
\]%
One obtains%
\begin{equation}
\mathcal{Q}=\frac{2^{s-1}q^{2s-1}}{16\pi l^{2s-1}}.  \label{Ch}
\end{equation}

Now, using Eqs. (\ref{Ent}-\ref{Ch}), the constant $C_{0}$ can be written
in terms of the thermodynamics quantities $T$, $\mathcal{S}$, $\Phi $, $%
\mathcal{E}$ and $\mathcal{Q}$ as
\[
C_{0}=16\pi T\mathcal{S}=16\pi\left(\frac{n+z-1}{n-1}\mathcal{E}-\mathcal{Q}\Phi\right),
\]%
for both the cases a and b, discussed in Sec. \ref{C0}. Thus, one
obtains

\begin{equation}
\mathcal{E}=\frac{(n-1)}{(n+z-1)}\left( T\mathcal{S}+\mathcal{Q}\Phi \right). \label{ETS}
\end{equation}
This is the generalization of the relation between the thermodynamic
quantities obtained in Ref. \cite{Therm}. For $z=1$, the energy density for AdS
spacetime will be obtained as
\[
\mathcal{E}=\frac{(n-1)}{n}(TS+\mathcal{Q}\Phi ),
\]%
which is the well-known Smarr formula.
\section{Conclusion}
In this paper, we consider asymptotically Lifshitz black branes in the presence of a massive electromagnetic
field and a massless nonlinear gauge field. The appropriate action of a charged black hole in
the grand-canonical and canonical ensembles are not the same \cite{Haw}. This is due to the fact
that the variation of the action will only give
the equations of motion if the variation is at a fixed gauge potential on the boundary, which
happens in the grand-canonical ensemble. But, in the canonical ensemble the charge is fixed, and, therefore,
the variation of the action will not give the equation of motion. The appropriate action in these ensembles is given
by Hawking and Ross for a linear electromagnetic field. Here, we generalized this action to the case
of power-law electromagnetic theory and introduce the appropriate action for both
the canonical and grand-canonical ensembles. We then present the condition on the parameters of the metric
for having nonextreme black brane solution. Also, we used the field equations to
find a conserved quantity along the $r$ coordinate. This
constant, which is the generalization of the constant introduced
in Ref. \cite{Pour}, has the role of connecting the metric parameters
at the horizon and at infinity. We used this conserved quantity to impose some conditions on
the nonlinear parameter $s$, the critical exponent, $z$ and the dimension of the spacetime.
We found that for $z>(n-1)/(2s-1)$ with $s>1/2$ and $z=n-1$ with $1/2<s<1$, the solutions are physical.
For these two cases, we calculate the energy density through the use of the counterterm method.
We also found the general thermodynamic relationship
for the energy density in terms of the extensive
thermodynamic quantities, entropy, and charge density, and
their intensive conjugate quantities, temperature and electric potential.
This result generalizes the well-known Smarr formula of AdS black holes.

We applied the counterterm method only to the case
of the $k=0$ solutions. It would be interesting to generalize this
method to the $k=\pm1$ cases. Further work in this area will involve considering
the thermodynamics of Lifshitz black holes of modified theory of gravity in the presence of
a nonlinear electromagnetic field. Also, investigating the effects of nonlinear
electromagnetic field on the stability of Lifshitz black branes in both the
canonical and grand-canonical ensembles is worth investigating.

\section*{Acknowledgements}
This work was supported by the Research Institute for
Astrophysics and Astronomy of Maragha.
\section{APPENDIX}
In this appendix we obtain the general form of the metric functions at large
$r$. Applying the small perturbation
\begin{eqnarray*}
&&f(r)=1+\epsilon f_{1}(r), \\
&&g(r)=1+\epsilon g_{1}(r), \\
&&h(r)=1+\epsilon h_{1}(r),
\end{eqnarray*}%
for the metric function at large $r$, the field equations up to the first
order in $\epsilon$ are

\begin{eqnarray*}
0 &=&2r^{2}h_{1}^{\prime \prime }+2(n+z)rh_{1}^{\prime }+zr\left(
g_{1}^{\prime }-f_{1}^{\prime }\right) +2(n-1)zg_{1}, \\
0 &=&\frac{1}{8}(2s-1)\left( \frac{2q^{2}}{l^{2}r^{2m}}\right)
^{s}+2(z-1)rh_{1}^{\prime }+(n-1)rg_{1}^{\prime }+\left[ z(z-1)+n(n-1)\right]
g_{1}-(z-1)(n+z-1)(f_{1}-2h_{1}), \\
0 &=&\frac{1}{8}(2s-1)\left( \frac{2q^{2}}{l^{2}r^{2m}}\right)
^{s}+2(z-1)rh_{1}^{\prime }+(n-1)rf_{1}^{\prime }+\left[
z(z-1)+n(n-1)+2(n-1)(z-1)\right] g_{1} \\
&&-(z-1)(z-n+1)(f_{1}-2h_{1}).
\end{eqnarray*}%
The solutions of the above system of linear equations are\newline
\textbf{A}. $z\neq (n-1)/(2s-1)$:\textbf{\ }
\begin{eqnarray*}
h_{1}(r) &=&-\frac{C_{1}}{r^{n+z-1}}-\frac{C_{2}}{r^{(n+z-1+\gamma )/2}}-%
\frac{C_{3}}{r^{(n+z-1-\gamma )/2}} \\
&&-\frac{(2s-1)^{4}(2q^{2})^{s}z[(n-1)s-1+z]}{2\mathcal{B}%
l^{2(s-1)}r^{2s(n-1)/(2s-1)}}, \\
f_{1}(r) &=&-\frac{C_{1}F_{1}}{r^{n+z-1}}-\frac{C_{2}F_{2}}{r^{(n+z-1+\gamma
)/2}}-\frac{C_{3}F_{3}}{r^{(n+z-1-\gamma )/2}} \\
&&+\frac{(2s-1)^{2}(2q^{2})^{s}%
\{-(2s-1)^{2}z^{2}-(2s-1)[(n-1)s^{2}-(n+1)s+1]z+(n-3)(n-1)s^{2}+(n-1)s\}}{%
\mathcal{B}l^{2(s-1)}r^{2s(n-1)/(2s-1)}}, \\
g_{1}(r) &=&-\frac{C_{1}G_{1}}{r^{n+z-1}}-\frac{C_{2}G_{2}}{r^{(n+z-1+\gamma
)/2}}-\frac{C_{3}G_{3}}{r^{(n+z-1-\gamma )/2}} \\
&&+\frac{s(n-1)(2s-1)^{2}[s(n-z+1)-2zs^{2}+z-1](2q^{2})^{s}}{\mathcal{B}%
l^{2(s-1)}r^{2s(n-1)/(2s-1)}},
\end{eqnarray*}%
\textbf{B}. $z=n-1$and $s\neq 1$:
\begin{eqnarray*}
f_{1}(r) &=&\frac{-(3n-4)C_{1}}{2(n-1)^{2}(2n-3)r^{2n-2}}-\frac{%
(2s-1)^{2}(2q^{2})^{s}[2s^{2}(n-1)+(2s-1)(n-2)]}{%
16(n-1)^{3}(s-1)s^{2}l^{2(s-1)}r^{2s(n-1)/(2s-1)}} \\
g_{1}(r) &=&-\frac{2(2n-3)(n-2)\ln (r)+(3n-4)}{2(n-1)(2n-3)^{2}r^{2n-2}}%
C_{1}-\frac{(n-2)C_{2}}{(n-1)(2n-3)r^{2n-2}} \\
&&-\frac{(2s-1)^{2}(2q^{2})^{s}[(n-1)(2s^{2}-1)-2s+1]}{%
16s(n-1)^{3}(s-1)^{2}l^{2(s-1)}r^{2s(n-1)/(2s-1)}} \\
h_{1}(r) &=&-\frac{2(n-1)\ln (r)+1}{4(n-1)^{2}r^{2n-2}}C_{1}-\frac{C_{2}}{%
2(n-1)r^{2n-2}}-\frac{(2s-1)^{4}(2q^{2})^{s}[(n-1)s+n-2]}{%
32s^{2}(n-1)^{3}(s-1)^{2}l^{2(s-1)}r^{2s(n-1)/(2s-1)}}
\end{eqnarray*}%
\textbf{C}. $z=(n-1)/(2s-1)$:
\begin{eqnarray}
h_{1}(r) &=&-\frac{C_{1}}{r^{n+z-1}}-\frac{C_{2}}{r^{(n+z-1+\gamma )/2}}-%
\frac{C_{3}}{r^{(n+z-1-\gamma )/2}}  \nonumber  \label{zns} \\
&&+\frac{(2q^{2})^{s}(2s-1)[2(n-1)s^{2}-(n+1)s+n]}{%
16s(s-1)(n-1)(n-2s)l^{2(s-1)}r^{2s(n-1)/(2s-1)}}\{\ln (r)+\frac{%
(2s-1)[(n-3)s^{2}+(n+2)s-n]}{2s(s-1)(n-1)(n-2s)}\},  \nonumber \\
f_{1}(r) &=&-\frac{C_{1}F_{1}}{r^{n+z-1}}-\frac{C_{2}F_{2}}{r^{(n+z-1+\gamma
)/2}}-\frac{C_{3}F_{3}}{r^{(n+z-1-\gamma )/2}}-\frac{(2s-1)(2q^{2})^{s}}{%
8s(n-1)l^{2(s-1)}r^{2s(n-1)/(2s-1)}}\{\ln (r)  \nonumber \\
&&+\frac{%
[4(n^{2}-4n+3)s^{5}+16(n-1)s^{4}-(5n^{2}+8n-13)s^{3}+2n(3n-1)s^{2}-2n(n+1)s+n^{2}](2s-1)%
}{2s(s-1)(n-1)(n-2s)\vartheta }\},  \nonumber \\
g_{1}(r) &=&-\frac{C_{1}G_{1}}{r^{n+z-1}}-\frac{C_{2}G_{2}}{r^{(n+z-1+\gamma
)/2}}-\frac{C_{3}G_{3}}{r^{(n+z-1-\gamma )/2}}  \nonumber \\
&&+\frac{(2s-1)(2q^{2})^{s}}{8(n-1)(s-1)l^{2(s-1)}r^{2s(n-1)/(2s-1)}}\left\{
\ln (r)+\frac{(2s-1)^{2}[2(n+1)s^{3}-(3n+7)s^{2}+2(2n+1)s-n]}{%
2(n-2s)(s-1)\vartheta }\right\} ,  \nonumber \\
&&.  \nonumber
\end{eqnarray}%
where
\begin{eqnarray*}
\mathcal{B} &=&4s(n-1)\{n-1+(1-2s)z\}\{[-4z^{2}+2(n+1)z-4(n-1)]s^{2} \\
&&+[4z^{2}-(3n+1)z+n(n+2)-3]s+(z-1)(n-z-1)\} \\
\vartheta  &=&2(n-1)s^{2}-(n+1)s+n, \\
\gamma  &=&\left\{ 9z^{2}-2(3n+1)z+n^{2}+6n-7\right\} ^{1/2}, \\
F_{1} &=&2\left( z-1\right) \left( z-n+1\right) {\mathcal{K}}^{-1}, \\
F_{2} &=&\left( \mathcal{F}_{1}-\mathcal{F}_{2}\right) \left\{ 8z\mathcal{K}%
\left[ (z-1)+2n+z-3\right] \right\} ^{-1}, \\
F_{3} &=&\left( \mathcal{F}_{1}+\mathcal{F}_{2}\right) \left\{ 8z\mathcal{K}%
\left[ (z-1)+2n+z-3\right] \right\} ^{-1}, \\
G_{1} &=&2\left( z-1\right) \left( n+z-1\right) {\mathcal{K}}^{-1}, \\
G_{2} &=&\left( \mathcal{G}_{1}+\mathcal{G}_{2}\right) \left\{ 8z\mathcal{K}%
\left[ (z-1)+2n+z-3\right] \right\} ^{-1}, \\
G_{3} &=&\left( \mathcal{G}_{1}-\mathcal{G}_{2}\right) \left\{ 8z\mathcal{K}%
\left[ (z-1)+2n+z-3\right] \right\} ^{-1}, \\
\mathcal{K} &=&(z-1)(n+z-1)+z(z-1)+n(n-1), \\
\mathcal{F}_{1} &=&8(z-1)[(z-1)(n+z-1)+z(z-1)+n(n-1)] \\
&&\times \lbrack (z-1)(n+3z-3)-2z^{2}+(n+3)z+n(n-2)-1], \\
\mathcal{F}_{2} &=&\gamma \lbrack n-1+(z-1)]\Big\{16(z-1)^{3} \\
&&+(17n-1)(z-1)^{2}+2(n+8)(n-1)(z-1)+n^{2}(n-1)-(n-1)\gamma ^{2}\Big\}, \\
\mathcal{G}_{1} &=&8(z-1)[2(z-1)-3z+3n-1][(z-1)(n+z-1)+z(z-1)+n(n-1)], \\
\mathcal{G}_{2} &=&\Big\{16(z-1)^{3}+(17n-1)(z-1)^{2}, \\
&&+2(n+8)(n-1)(z-1)+n^{2}(n-1)\Big\}\gamma -(n-1)\gamma ^{3}.
\end{eqnarray*}


\begin{thebibliography}{99}
\bibitem{Alt} A. Altland and B. Simons, Condensed Matter Field Theory
(Cambridge University Press, Cambridge, England, 2006).

\bibitem{Sach} S. Sachdev, Quantum Phase Transitions (Cambridge University
Press, Cambridge, England 1999).

\bibitem{Hertz} J. A. Hertz, Phys. Rev. B \textbf{14}, 1165 (1976).

\bibitem{Kach} Shamit Kachru, Xiao Liu, and Michael Mulligan, Phys. Rev. D
\textbf{78}, 106005 (2008).

\bibitem{Mal} J. Maldacena, Adv. Th. Math. Phys. \textbf{2}, 231 (1998); O.
Aharony, S. S. Gubser, J. M. Maldacena, H. Ooguri, and Y. Oz, Phys. Rep.
\textbf{323}, 183 (2000).

\bibitem{Hart} S. A. Hartnoll, Classical Quantum Gravity \textbf{26}, 224002
(2009); S. A. Hartnoll, C. P. Herzog and G.T. Horowitz, Phys. Rev. Lett.
\textbf{101}, 031601 (2008); S. A. Hartnoll and P. K. Kovtun, Phys. Rev. D
\textbf{76}, 066001 (2007); C. P. Herzog, J. Phys. A \textbf{42}, 343001
(2009).

\bibitem{omid} S. F. Ross and O. Saremi, J. High Energy Phys. \textbf{09},
009 (2009).

\bibitem{Pang} D. W. Pang, J. High Energy Phys. \textbf{01}, 116 (2010);

\bibitem{Exact} E. J. Brynjolfsson, U. H. Danielsson, L. Thorlacius and T.
Zingg, J. Phys. A \textbf{43}, 065401 (2010); K. Balasubramanian and J.
McGreevy, Phys. Rev. D \textbf{80}, 104039 (2009); E. Ayon-Beato, A.
Garbarz, G. Giribet and M. Hassaine, Phys. Rev. D \textbf{80}, 104029
(2009); R. G. Cai, Y. Liu and Y. W. Sun, J. High Energy Phys. \textbf{10},
080 (2009); E. Ayon-Beato, A. Garbarz, G. Giribet and M. Hassaine, J. High
Energy Phys. \textbf{04}, 030 (2010); W. Chemissany and J. Hartong,
Classical Quantum Gravity \textbf{28}, 195011 (2001); H. Maeda and G. Giribet, J. High Energy
Phys. \textbf{07}, 015 (2011); J. Tarrio and S. Vandoren, J. High Energy Phys. \textbf{09}, 017 (2011).

\bibitem{Num} U. H. Danielsson, L. Thorlacius, J. High Energy Phys. \textbf{%
03}, 070 (2009); R. B. Mann, J. High Energy Phys. \textbf{06}, 075 (2009);
G. Bertoldi, B. A. Burrington, A. W. Peet, Phys. Rev. D \textbf{80} (2009)
126003; M. H. Dehghani and R. B. Mann, J. High Energy Phys. \textbf{07}, 019
(2010); W. G. Brenna, M. H. Dehghani and R. B. Mann, Phys. Rev. D \textbf{84}%
, 024012 (2011); I. Amado and A. F. Faedo, J. High Energy Phys. \textbf{07}, 004 (2011).

\bibitem{Therm} G. Bertoldi, B. A. Burrington and A. W. Peet, Phys. Rev. D
\textbf{80}, 126004 (2009).

\bibitem{MH} M. H. Dehghani and R. B. Mann, Phys. Rev. D \textbf{82}, 064019
(2010); M. H. Dehghani and Sh. Asnafi, Phys. Rev. D \textbf{84}, 064038
(2011).

\bibitem{Pour} M. H. Dehghani, R. B. Mann and R. Pourhasan, Phys. Rev. D
\textbf{84}, 046002 (2011).

\bibitem{Gross} D.J. Gross and J.H. Sloan, Nucl. Phys. B \textbf{291}, 41
(1987).

\bibitem{Mart} M. Hassaine and C. Martinez, Classical Quantum Gravity \textbf{25%
}, 195023 (2008).

\bibitem{Hendi} H. Maeda, M. Hassaine and C. Martinez, Phys. Rev. D \textbf{%
79}, 044012 (2009); S. H. Hendi, Phys. Lett. B \textbf{678}, 438 (2009); S.
H. Hendi, Classical Quantum Gravity \textbf{26}, 225014 (2009); S. Habib
Mazharimousavi and M. Halilsoy, Phys. Lett. B \textbf{681}, 190 (2009); A.
Bazrafshan, M. H. Dehghani, M. Ghanaatian, Phys. Rev. D \textbf{86}, 104043
(2012).

\bibitem{Haw} S. W. Hawking and S. F. Ross, Phys. Rev. D \textbf{52}, 5865
(1995).

\bibitem{Can} M. M. Caldarelli, G. Cognola, and D. Klemm, Classical Quantum
Gravity \textbf{17}, 339 (2000); M. H. Dehghani and A. Khodam-Mohammadi, Phys. Rev. D \textbf{67},
084006 (2003).

\end{thebibliography}
\end{document}